\documentclass[aps]{revtex4}

\usepackage{amsmath}
\usepackage{graphicx}

\begin{document}

\title{Effective long-time phase dynamics of limit-cycle oscillators\\driven by weak colored noise}

\author{Hiroya Nakao$^{1,2}$}
\author{Jun-nosuke Teramae$^{3,4}$}
\author{Denis S. Goldobin$^{5,6}$}
\author{Yoshiki Kuramoto$^{7,8}$}

\affiliation{$^{1}$Department of Physics, Kyoto University, Kyoto 606-8502, Japan}
\affiliation{$^{2}$JST, CREST, Kyoto 606-8502, Japan}
\affiliation{$^{3}$RIKEN Brain Science Institute, Wako 2-1, Saitama, Japan}
\affiliation{$^{4}$PRESTO, Japan Science and Technology Agency (JST), 4-1-8 Honcho Kawaguchi, Saitama 332-0012, Japan}
\affiliation{$^{5}$Institute of the Continuous Media Mechanics, UB RAS, Perm 614013, Russia}
\affiliation{$^{6}$Department of Mathematics, University of Leicester, Leicester LE1 7RH, UK}
\affiliation{$^{7}$Research~Institute~for~Mathematical~Science,~Kyoto~University,~Kyoto~606-8502,~Japan}
\affiliation{$^{8}$Institute~for~Integrated~Cell-Material~Sciences,~Kyoto~University,~Kyoto~606-8501,~Japan}
\date{\today}

\begin{abstract}
An effective white-noise Langevin equation is derived that describes long-time phase dynamics of a limit-cycle oscillator subjected to weak stationary colored noise.
Effective drift and diffusion coefficients are given in terms of the phase sensitivity of the oscillator and the correlation function of the noise, and are explicitly calculated for oscillators with sinusoidal phase sensitivity functions driven by two typical colored Gaussian processes.
The results are verified by numerical simulations using several types of stochastic or chaotic noise.
The drift and diffusion coefficients of oscillators driven by chaotic noise exhibit anomalous dependence on the oscillator frequency, reflecting the peculiar power spectrum of the chaotic noise.
\end{abstract}

 
\maketitle

{\bf
Limit-cycle oscillators are used to model a variety of rhythmic processes in nature.  When a limit cycle is subjected to noise, the frequency of oscillations changes and the oscillation phase tends to diffuse.  These effects are quantified by drift and diffusion coefficients and are important in understanding the long-time behavior of noisy oscillators.  Here, we derive their analytical expressions in terms of the phase sensitivity function of the limit cycle and the correlation function of the noise, and verify them by numerical simulations using several types of stochastic or chaotic noise.  Our formulation will provide a simple and general way to analyze the long-time dynamics of limit-cycle oscillators driven by arbitrary weak and smooth colored noise.
}

\section{Introduction}

Nonlinear oscillations are ubiquitously observed in nature and various models of limit-cycle oscillators with stable periodic dynamics have been used to describe them~\cite{Winfree0,Winfree}.  Some of the many examples are oscillatory chemical reactions, cardiac cells, spiking neurons, circadian rhythms, frog calls, passively walking robots, and pedestrians on a bridge~\cite{Winfree0,Winfree,Kuramoto,Pikovsky,Koch,Aihara,McGeer,Ott}.
A powerful technique for analyzing weakly perturbed limit-cycle oscillators is {\it phase reduction}~\cite{Winfree0,Winfree,Kuramoto,Pikovsky}, which approximately describes a limit-cycle oscillator possessing multi-dimensional state variables by a single phase variable.  The resulting one-dimensional phase equation is solely specified by the {\it frequency} and the {\it phase sensitivity} derived from the original limit-cycle oscillator, which greatly facilitates analytical treatments~\cite{Winfree,Kuramoto}.  The diverse dynamics of limit-cycle oscillators driven by external forcing or coupled by mutual interactions have been analyzed using phase reduction~\cite{Rinzel,Hansel,Kopell,Ermentrout1}.

Since all systems in nature are subjected to fluctuations, it is essential to incorporate the effect of noise into the dynamics of limit-cycle oscillators. It is well documented that noise can induce nontrivial dynamics in oscillator systems~\cite{Winfree,Kuramoto,Shinomoto,Pikovsky,Haken,Rappel,Kurrer,Kiss,Kawamura,Gil}.
Synchronization among noninteracting limit-cycle oscillators induced by common or shared noisy forcing is a prominent example and has garnered considerable interest in connection with the reproducibility of lasers and electronic circuits~\cite{Uchida,Yoshida,Arai-Nakao,Nagai-Nakao2}, synchrony of spiking neurons~\cite{Mainen-Sejnowski,Binder-Powers,Galan,Tateno-Robinson}, and large-scale correlated fluctuations in ecosystems~\cite{Moran,Koenig,Ranta,Royama}.
Phase reduction methods have been extensively used to analyze this phenomenon for limit-cycle oscillators driven by various types of common noise (weak Gaussian noise~\cite{Pakdaman,Teramae-Dan,Goldobin-Pikovsky,Nakao-Arai-Kawamura,Galan2}, Poisson random impulses~\cite{Nakao-Arai}, and others~\cite{Nagai-Nakao}).
As discussed in~\cite{Yoshimura-Arai,Nakao-Teramae-Ermentrout,Teramae-Nakao-Ermentrout}, phase reduction methods should be applied to noise-driven limit-cycle oscillators with care, in order to properly consider the effect of amplitude fluctuations; this is in contrast to the case with smooth forcing, where phase reduction can be performed without ambiguity.  In~\cite{Goldobin2}, a general attempt is made to derive a phase equation, which explicitly takes into account the effect of amplitude dynamics, for a wide class of noise.

In this paper, we analyze noisy limit cycles from an alternative viewpoint, namely, their long-time stochastic phase dynamics.
In particular, we will derive an effective white-noise Langevin equation that describes the coarse-grained phase dynamics of a limit-cycle oscillator driven weakly by sufficiently smooth stationary colored noise.
In deterministic systems of limit cycles, e.g., in the synchronization process of coupled oscillators, long-time phase dynamics dominate the entire system behavior.  Similarly, the long-time behaviors of limit cycles should play crucial roles in stochastic oscillator systems and methods for treating them should be developed.
Note that the extraction of effective dynamics of slow modes has been a classical topic in the theory of stochastic processes (and not just in the context of nonlinear oscillators) and various methods such as the use of projection operators and multiscale expansion have been developed~\cite{Zwanzig,Doering,Reimann,Just,Pavliotis,Stemler,Stuart}.

In the present case, the amplitude effect of the oscillator does not play a significant role at the lowest order approximation because its decay is much more rapid than the phase dynamics~\cite{Teramae-Nakao-Ermentrout,Goldobin2}, and hence the conventional phase equation holds.
We focus on how to obtain drift and diffusion coefficients by specifying the effective Langevin equation that gives the long-time phase dynamics of the limit cycle.
To this end, we develop a simple theory based on the Kramers-Moyal expansion~\cite{Risken,Gardiner}, which gives the effective drift and diffusion coefficients in terms of the phase sensitivity of the limit cycle and the correlation function of the  applied noise.
Using several types of phase sensitivity functions and noisy signals, we demonstrate how the effective drift and diffusion coefficients depend on the characteristics of the driving noise.

\section{Theory}

In this section, we derive an effective Langevin equation describing the long-time phase dynamics of a limit-cycle oscillator driven by sufficiently weak and smooth stationary colored noise.  We introduce a timescale at which our effective description holds, and calculate the drift and diffusion coefficients from the phase sensitivity function of the oscillator and the correlation function of the noise.

\subsection{Model}

We consider a limit-cycle oscillator driven by noise,
\begin{align}
  \dot{\bf X}(t) = {\bf F}({\bf X}) + \epsilon {\boldsymbol \xi}(t),
  \label{ODE}
\end{align}
where the vector ${\bf X}(t)$ is the state of the oscillator at time $t$, ${\bf F}({\bf X})$ is the intrinsic dynamics of the oscillator, ${\boldsymbol \xi}(t)$ is the noise, and $\epsilon$ is a small parameter representing noise intensity.  We assume that Eq.~(\ref{ODE}) has a stable limit-cycle solution ${\bf X}_0(t+T) = {\bf X}_0(t)$ with period $T$ when the noise is absent ($\epsilon = 0$).
The noise is assumed to be smooth, so that ordinary rules of differential calculus apply for the variable ${\bf X}(t)$.  More explicitly, we consider the cases in which the noise is given by some time-integrated process of (i) stochastic differential equations with Gaussian white noise or (ii) ordinary differential equations with chaotic dynamics.

When the noise is sufficiently small ($\epsilon \ll 1$), the oscillator state can approximately be described using only its phase~\cite{Winfree0,Winfree,Kuramoto}.  We first introduce a phase $\phi \in [0, 2\pi)$ on the unperturbed limit-cycle orbit ${\bf X}_0(t)$ that increases with a constant rate (frequency) $\omega = 2\pi / T$ as $\dot{\phi}(t) = \omega$.  This phase $\phi$ can then be extended as a phase field $\phi({\bf X})$ around ${\bf X}_0(t)$ in such a way that $\dot{\phi}(t) = \nabla_{\bf X} \phi({\bf X}) \cdot {\bf F}({\bf X}) = \omega$ holds constantly.  
The dynamics of $\phi$ at the lowest order in $\epsilon$ are given by
\begin{align}
  \dot{\phi}(t) = \omega + \epsilon Z(\phi(t)) \xi(t),
  \label{Colored}
\end{align}
where, for simplicity, it is assumed that the noise ${\boldsymbol \xi}(t)$ is given only to a single vector component $X_{i}$ ($i=1, \cdots, N)$ of ${\bf X}$, and we denote its intensity by a scalar function $\xi(t)$.
The $2\pi$ periodic function
\begin{align}
Z(\phi) = \left. \frac{ {\partial \phi({\bf X})} }{ {\partial X_{i}} } \right|_{{\bf X} = {\bf X}_0(\phi)}
\end{align}
is called the phase sensitivity~\cite{Winfree0,Winfree,Kuramoto}, representing a linear response coefficient of the phase $\phi$ to tiny perturbations applied to the vector component $X_{i}$ of the oscillator.  Extension to general vector noise is straightforward.

We assume that $\xi(t)$ is a zero-mean stationary random process generated by some noise source, which is smooth, temporally correlated, and generally non-Gaussian, with a two-point correlation function $C(t)$, namely,
\begin{align}
  \langle \xi(t) \rangle = 0, \;\;\; \langle \xi(t) \xi(0) \rangle = C(t),
\end{align}
where $\langle \cdots \rangle$ represents the ensemble average.  We further assume that the correlation function decays with a characteristic time $\tau_{c}$ as $|C(t)| = O( e^{-|t| / \tau_{c}} )$.

\subsection{Separation of timescales}

Our goal is to derive an effective Langevin equation with Gaussian white noise that approximates the long-time dynamics of Eq.~(\ref{Colored}).  To proceed, we introduce a new slow phase variable by $ \psi(t) = \phi(t) - \omega t $ and rewrite Eq.~(\ref{Colored}) as
\begin{align}
  \dot{\psi}(t) = \epsilon Z(\omega t + \psi(t)) \xi(t).
	\label{Slow}
\end{align}
Let $\tau_{P} = \epsilon^{-1}$ represent a timescale of the slow phase dynamics of $\psi$, where $\tau_{p}$ is much larger than the characteristic decay time $\tau_{c}$ of the noise correlation $C(t)$, i.e., $\tau_{c} \ll \tau_{P}$.

For sufficiently small $\epsilon$, we can introduce an intermediate timescale $\tau$, which is sufficiently longer than the noise correlation time, $ \tau_{c} \ll \tau$, but still the slow phase $\psi$ does not change significantly within $[t, t + \tau]$, namely,
\begin{align}
	| \psi(t + \tau) - \psi(t) | \ll 1.
\end{align}
 This condition implies $ \tau \ll \tau_P = \epsilon^{-1} $ because $| \psi(t + \tau) - \psi(t) |  = O( \epsilon \tau )$ for bounded $Z(\phi)$ and $\xi(t)$. 

Thus, we have three distinct timescales in our problem, which satisfy
\begin{align}
	\tau_{c} \ll \tau \ll \tau_{P}.
\end{align}
The separation of timescales allows us to derive an effective Gaussian white stochastic process from Eq.~(\ref{Slow}) at the long timescale $\tau_{P}$ that describes the slow dynamics of $\psi$ by renormalizing fast fluctuations of the noise $\xi(t)$ at the short timescale $\tau_{c}$ into effective drift and diffusion coefficients.

\subsection{Effective Langevin equation}

We use a simple Fokker-Planck approximation to the Kramers-Moyal equation~\cite{Risken} describing the dynamics of a probability density function (PDF) $P(\psi, t)$ of $\psi$ corresponding to Eq.~(\ref{Slow}), using the periodicity of the phase sensitivity function $Z(\phi)$.
To this end, we calculate the first- and second-order moments of the slow phase dynamics of $\psi$ during $[t, t+\tau]$,
\begin{align}
M_1(\psi, t; \tau) = \langle \psi(t + \tau) - \psi(t) \rangle,
\;\;\;
M_2(\psi, t; \tau) = \langle ( \psi(t + \tau) - \psi(t) )^2 \rangle,
\end{align}
where the ensemble average $\langle \cdots \rangle$ is taken over noise realizations with fixed $\psi$ and $t$.  Effective drift and diffusion coefficients $v(\psi, t)$ and $D(\psi, t)$, respectively, of the approximate Fokker-Planck equation are obtained from these moments at the long timescale $\tau_{P} ( \gg \tau )$ by ignoring fast fluctuations of the noise at the short timescale $\tau_{c} ( \ll \tau )$.  That is, we regard $\tau$ as a small parameter, retain only the $O(\tau)$ term, and then formally take the $\tau \to 0$ limit;  this yields the effective drift and diffusion coefficients
\begin{align}
  v(\psi, t) = \lim_{\tau \to 0} \frac{M_1(\psi, t; \tau)}{\tau},
  \;\;\;
  D(\psi, t) = \lim_{\tau \to 0} \frac{M_2(\psi, t; \tau)}{\tau},
  \label{EfCf}
\end{align}
where $v(\psi, t)$ and $D(\psi, t)$ will turn out to be constants independent of $\psi$ and $t$.  The resulting approximate Fokker-Planck equation
\begin{align}
  \frac{\partial}{\partial t} P(\psi, t) = - \frac{\partial}{\partial \psi} [ v P(\psi, t) ] + \frac{D}{2} \frac{\partial^2}{\partial \psi^2} P(\psi, t)
  \label{SlowFPE}
\end{align}
corresponds to an effective Langevin equation,
\begin{align}
  \dot{\psi}(t) = v + \sqrt{D} \eta(t),
  \label{EffLan}
\end{align}
where $\eta(t)$ is Gaussian white noise satisfying $\langle \eta(t) \rangle = 0$ and $\langle \eta(t) \eta(s) \rangle = \delta(t-s)$.
Considering the original phase variable $\phi(t)$, the effective Langevin equation will be given by
\begin{align}
  \dot\phi(t) = \omega + v + \sqrt{D} \eta(t).
\end{align}
Thus, the effective drift coefficient $v$ gives a noise-induced correction to the raw oscillator frequency $\omega$.  The diffusion coefficient $D$ gives the effective intensity of the noise.
Note that we make no assumption on the oscillator frequency $\omega$ but assume only that $\tau_{c} \ll \tau \ll \tau_{P} = \epsilon^{-1}$, which can always be satisfied for sufficiently small $\epsilon$.  In other words, we can always find a scaling region where the above effective Langevin equation is valid, as long as the noise is sufficiently weak.

\subsection{Drift and diffusion coefficients}

To calculate the moments $M_{1}(\psi, t ; \tau)$ and $M_{2}(\psi, t ; \tau)$ explicitly, we expand the phase sensitivity function $Z(\phi)$ estimated at $\phi(t_{1}) = \omega t_{1} + \psi(t_{1})$ as
\begin{align}
  Z(\omega t_{1} + \psi(t_{1})) 
  &= Z(\omega t_{1} + \psi(t) + \psi(t_{1}) - \psi(t)) \cr
  &= Z(\omega t_{1} + \psi(t)) + Z'(\omega t_{1} + \psi(t)) \{ \psi(t_{1}) - \psi(t) \}
  + O( \{ \psi(t_{1}) - \psi(t) \}^{2} ),
\end{align}
where $Z'(\phi) = d Z(\phi) / d\phi$, and we integrate Eq.~(\ref{Slow}) as
\begin{align}
  \psi(t + \tau) - \psi(t)
  &= \epsilon \int_{t}^{t + \tau} dt_{1} Z(\omega t_{1} + \psi(t)) \xi(t_{1}) \cr
  &+ \epsilon \int_{t}^{t + \tau} dt_{1} Z'(\omega t_{1} + \psi(t))  \{ \psi(t_{1}) - \psi(t) \} \xi(t_{1})
  +  O(\epsilon \{ \psi(t_{1}) - \psi(t) \} ).
  \;\;\;\;\;\;
\end{align}
By iterative substitution, namely, by inserting $\psi(t_{1}) - \psi(t) = \epsilon \int_{t}^{t_{1}} dt_{2} Z'(\omega t_{2} + \psi(t)) \xi(t_{2}) + O(\epsilon^{2})$ obtained from the above equation into its second term, we obtain
\begin{align}
  \psi(t + \tau) - \psi(t)
  &= \epsilon \int_{t}^{t + \tau} dt_{1} Z(\omega t_{1} + \psi(t)) \xi(t_{1}) \cr
  &+ \epsilon^2 \int_{t}^{t + \tau} dt_{1} \int_{t}^{t_{1}}
  dt_{2} Z'(\omega t_{1} + \psi(t)) Z(\omega t_{2} + \psi(t)) \xi(t_{1}) \xi(t_{2}) 
  + O(\epsilon^3, \tau^{2}),\;\;\;
  \label{Expansion}
\end{align}
where we have used the fact that $| \psi(t + \tau) - \psi(t) | = O( \epsilon \tau )$.  Taking the ensemble average of this expression over the noise, the moments $M_1(\psi, t; \tau)$ and $M_2(\psi, t; \tau)$ can be calculated up to $O(\epsilon^{3}, \tau^{2})$ as
\begin{align}
  M_1(\psi, t ; \tau)
  &=
  \epsilon^2
  \left[ \int_{t}^{t + \tau} dt_{1} \int_{t}^{t_{1}}
  dt_{2} Z'(\omega t_{1} + \psi) Z(\omega t_{2} + \psi) C(t - t_{1})
  \right]
    + O(\epsilon^{3}, \tau^{2}),
      \label{M10}
\end{align}
\begin{align}
  M_2(\psi, t ; \tau)
  &=
  \epsilon^{2}
  \left[ \int_{t}^{t + \tau} dt_{1} \int_{t}^{t + \tau}
  dt_{2} Z(\omega t_{1} + \psi) Z(\omega t_{2} + \psi) C(t - t_{1}) \right]
  + O(\epsilon^{3}, \tau^{2}).
      \label{M20}
\end{align}

Now we take $\tau$ as an integer multiple of $T = 2\pi / \omega$, namely,
\begin{align}
\tau = n T = 2 n \pi/ \omega
\end{align}
with some appropriate integer $n \ (=1, 2, \cdots)$ such that $ \tau_{c} \ll \tau = n T \ll \tau_{P} = \epsilon^{-1}$ is satisfied (we may simply set $n=1$ if $\tau_{c} \ll T$).
Using the periodicity of the phase sensitivity function, the moments can be written as 
\begin{align}
  M_1(\psi, t ; \tau)
  &=
  \epsilon^2 \tau
  \left[ \frac{1}{2 \pi} 
  \int_{0}^{\infty} ds C(s)
  \int_{0}^{2 \pi} d\theta Z'(\theta) Z(\theta - \omega s)
  \right]
  + O(\epsilon^{3}, \tau^{2}),
  \label{M1}
\end{align}
\begin{align}
  M_2(\psi, t ; \tau)
  &=
  \epsilon^2 \tau \left[ \frac{ 1 }{2 \pi} 
  \int_{-\infty}^{\infty} ds C(s)
  \int_{0}^{2 \pi} d\theta Z(\theta) Z(\theta - \omega s) \right]
  + O(\epsilon^{3}, \tau^{2}),
  \label{M2}
\end{align}
both of which turn out to be constants (see Appendix for calculations).

From Eq.~(\ref{EfCf}), the effective drift and diffusion coefficients are obtained as
\begin{align}
  v =
\epsilon^2 \left[ \frac{1}{2 \pi}
  \int_{0}^{\infty} ds C(s)
  \int_{0}^{2 \pi} d\theta Z'(\theta) Z(\theta - \omega s) \right],
  \label{Effv}
\end{align}
and
\begin{align}
  D = \epsilon^2 \left[ \frac{1}{2 \pi}
  \int_{-\infty}^{\infty} ds C(s)
  \int_{0}^{2 \pi} d\theta Z(\theta) Z(\theta - \omega s) \right].
  \label{EffD}
\end{align}
The colored noise gives constant contributions of $O(\epsilon^{2})$ to both $v$ and $D$.
In~\cite{Galan4} and~\cite{Ly}, similar weak-noise expansion methods for the phase dynamics of noise-driven limit cycle oscillators are used to estimate their Lyapunov exponent (Eq.~(8) in~\cite{Galan4}, which generalizes the result in~\cite{Teramae-Dan} for Gaussian noise) or the variance of periods.

\subsection{Fourier representation}

The effective drift and diffusion coefficients can be expressed concisely using the Fourier representation of the phase sensitivity function,
\begin{align}
  Z(\theta) = \sum_{\ell = -\infty}^{\infty} \tilde{Z}_{\ell} e^{i \ell \theta},
\end{align}
as well as the power spectrum of the noise $\xi(t)$,
\begin{align}
  I(\Omega) = \int_{-\infty}^{\infty} C(t) e^{i \Omega t} dt.
\end{align}
Because $\xi(t)$ is stationary, the correlation function satisfies $C(t) = C(-t)$, so that the power spectrum $I(\Omega)$ can be expressed as 
\begin{align}
  I(\Omega) = 2 \int_{0}^{\infty} C(t) \cos(\Omega t) dt = 2 \mbox{Re} \chi(\Omega),
\end{align}
where
\begin{align}
  \chi(\Omega) = \int_{0}^{\infty} C(t) e^{i \Omega t} dt
\end{align}
is the Fourier-Laplace transform of the correlation function. 
Since $\chi(\Omega)$ is analytic in the upper-half of the complex plane ($\mbox{Re}\ \Omega > 0$), we can express its imaginary part using the Kramers-Kronig relation as
\begin{align}
  \mbox{Im} \chi(\Omega)
  = \frac{1}{\pi} \mbox{P.V.}\int_{-\infty}^{\infty}
  \frac{ \mbox{Re} \chi(z) }{ \Omega - z } dz
  = \frac{1}{2} \hat{I}(\Omega),
\end{align}
where 
\begin{align}
\hat{I}(\Omega) = \frac{1}{\pi} \mbox{P.V.}\int_{-\infty}^{\infty}
  \frac{ I(z) }{ \Omega - z } dz
\end{align}
is a Hilbert transform of the power spectrum $I(\Omega)$~(see e.g.,~\cite{Arfken}).  Inserting these equations into Eqs.~(\ref{Effv})~and~(\ref{EffD}), the drift and diffusion coefficients $v$ and $D$ can be expressed as
\begin{align}
  v = 
  \frac{\epsilon^2}{2} \sum_{\ell=-\infty}^{\infty} (i \ell) |\tilde{Z}_{\ell}|^2 \left\{ I(\omega \ell ) + i \hat{I}( \omega \ell ) \right\},
  \label{v_ft}
\end{align}
and
\begin{align}
  D = \epsilon^2 \sum_{\ell=-\infty}^{\infty} |\tilde{Z}_{\ell}|^2 I(\omega \ell).
  \label{D_ft}
\end{align}
Thus, $v$ and $D$ can be calculated from the power spectrum $I(\Omega)$ and its Hilbert transform $\hat{I}(\Omega)$.  This is convenient because the phase sensitivity $Z(\phi)$ often contains only lower harmonic components.

\section{Examples}  

In this section, we numerically verify the accuracy of the effective white-noise phase Langevin equation~(\ref{EffLan}) for several types of colored noise generated by stochastic processes and deterministic chaotic systems.  We compare the effective drift and diffusion coefficients given in Eqs.~(\ref{Effv}) and (\ref{EffD}), respectively, with those obtained by direct numerical simulations of the original phase model, Eq.~(\ref{Slow}).

\begin{figure}[tbp]
  \begin{center}
    \includegraphics[width=0.55\hsize,clip]{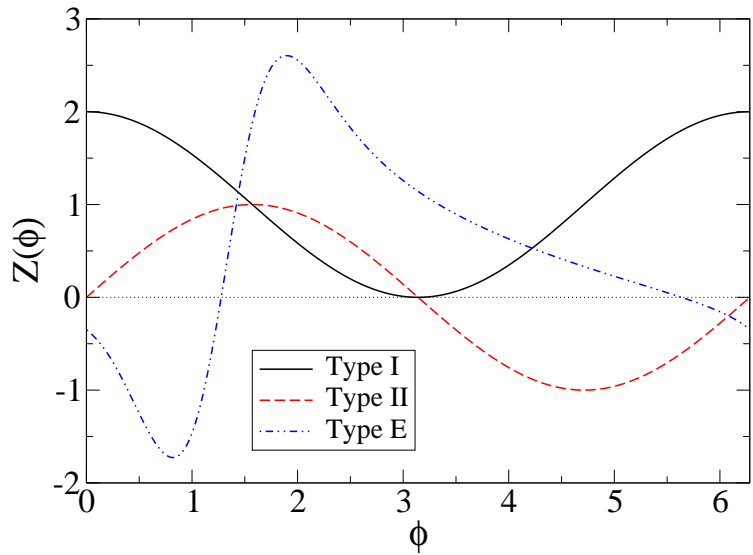}
    \caption{Type-I sinusoidal phase sensitivity $Z_{I}(\phi)$, Type-II sinusoidal phase sensitivity $Z_{II}(\phi)$, and Type-E phase sensitivity $Z_{E}(\phi)$ obtained from a Morris-Lecar spiking neuron model.}
    \label{fig-Z}
  \end{center}
\end{figure}

\subsection{Phase sensitivity functions}

We consider the following examples of phase sensitivity functions (see Fig.~\ref{fig-Z}):
\begin{enumerate}

\item Type-I sinusoidal function with only a positive lobe, corresponding to limit cycles near saddle-node bifurcation~\cite{Ermentrout1,Ermentrout2,Rinzel},
\begin{align}
  Z_{I}(\phi) = 1 + \cos \phi.
 \label{typeI}
\end{align}
\item Type-II sinusoidal function with positive and negative lobes, corresponding to limit cycles near Hopf bifurcation~\cite{Kuramoto},
\begin{align}
  Z_{II}(\phi) = \sin \phi
 \label{typeII}.
\end{align}
\item Phase sensitivity function $Z_{E}(\phi)$ of the Morris-Lecar neuron model near homoclinic bifurcation (see Appendix), where the function is not simply sinusoidal but contains higher-order harmonics.  It can be calculated numerically by the adjoint method~\cite{Kopell,Holmes}.  We refer to this function as Type-E.
\end{enumerate}
The first two functions $Z_{I, II}(\phi)$ are generic in the sense that they can be derived analytically from the normal forms of limit-cycle oscillators near the respective bifurcation points by appropriate coordinate transformations~\cite{Ermentrout1,Holmes}.  The third function $Z_{E}(\phi)$ is model-dependent, but is a typical example of the phase sensitivity near a homoclinic bifurcation point.  $Z_{E}(\phi)$ tends to be dominated by an exponentially decaying part resulting from the linear dynamics near a saddle point as the bifurcation point is approached, and therefore a simple exponential function with a discontinuity is proposed as a generic form of the phase sensitivity in~\cite{Holmes}.  We do not however use this form to avoid unnatural effects of the artificial discontinuity.

\begin{figure}[tbp]
  \begin{center}
    \includegraphics[width=1.0\hsize,clip]{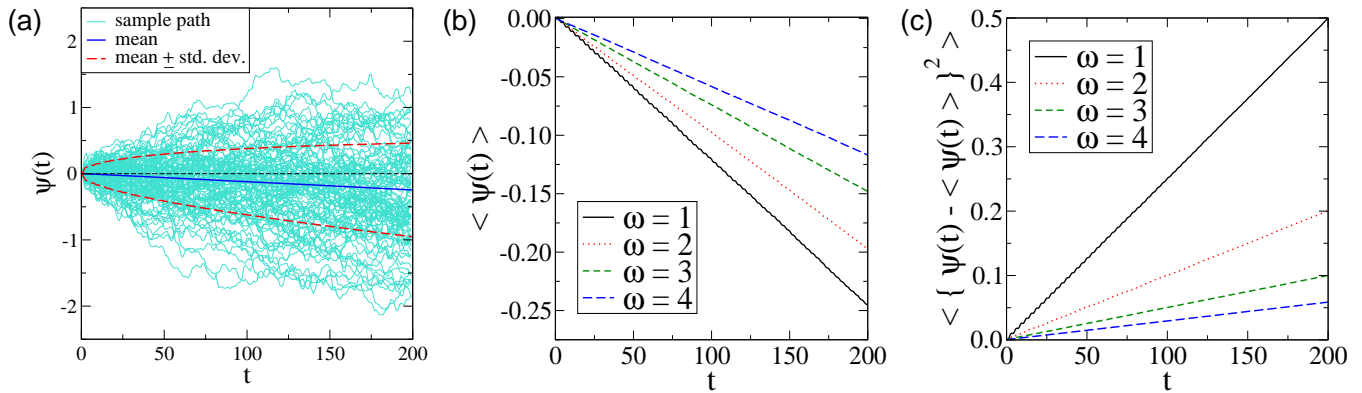}
    \caption{(a): Sample paths of the slow phase $\psi(t)$ driven by the Ornstein-Uhlenbeck noise ($100$ realizations).  Oscillator frequency $\omega = 1$ and the phase sensitivity function is $Z_{II}(\phi)$.  Noise correlation time $\tau_{c} = 1$ and noise intensity $\epsilon = 0.1$.  Solid lines represent the mean value and broken curves indicate the mean value $\pm$ the standard deviation. (b) and (c): Temporal growth of the mean (b) and the variance (c) of the phase $\psi(t)$ averaged over $200,000$ realizations for $\omega = 1, 2, 3$, and $4$. The other parameters are the same as in (a).}
    \label{fig1}
  \end{center}
\end{figure}

\subsection{Measuring the coefficients}

We estimate the effective drift and diffusion coefficients $v$ and $D$ by direct numerical simulations of Eq.~(\ref{Slow}) and compare them with the respective theoretical values, Eqs.~(\ref{Effv}) and (\ref{EffD}).  The solution to the effective Fokker-Planck equation~(\ref{SlowFPE}) from a delta-peaked initial condition $P(\psi, 0) = \delta(\psi)$ is simply a Gaussian wave packet,
\begin{align}
  P(\psi, t) = \frac{1}{\sqrt{2 \pi D t}} \exp\left[ -\frac{(\psi - v t)^2}{2 D t} \right],
\end{align}
whose moments are given by
\begin{align}
  \langle \psi(t) \rangle = v t, \;\;\;
  \langle \{ \psi(t) - \langle \psi(t) \rangle \}^2 \rangle = D t.
\end{align}
Thus, we can measure $v$ and $D$ from slopes of the mean and the variance of the phase plotted as functions of $t$.

For example, Fig.~\ref{fig1}(a) displays typical sample paths of Eq.~(\ref{Slow}) with the Type-II function $Z_{II}(\phi)$.  The evolution of the slow phase $\psi(t) = \phi(t) - \omega t$ is plotted for $100$ realizations of the OU noise (explained below).  The broken line represents the mean path averaged over $200,000$ realizations, which shows negative drift induced by the finite correlation time $\tau_{c} = 1$ of the noise.
Figures~\ref{fig1}(b) and (c) display the mean and the variance of the oscillator phase, respectively, averaged over $200,000$ realizations for differing values of $\omega$ and for the Type-II $Z(\phi)$, all of which clearly show linear dependence on time $t$, whose slopes yield $v$ and $D$.

\subsection{Ornstein-Uhlenbeck noise}

We first consider the case in which the colored noise $\xi(t)$ obeys the OU process,
\begin{align}
\dot{\xi}(t) = - \frac{1}{\tau_{c}} \xi + \frac{1}{\tau_{c} } \eta(t),
\end{align}
where $\eta(t)$ is zero-mean Gaussian white noise whose correlation function is given by $\langle \eta(t) \eta(s) \rangle = \delta(t-s)$.  This OU process generates colored Gaussian noise $\xi(t)$ with a stationary PDF
\begin{align}
P(\xi) = \left( \frac{\tau_{c}}{\pi} \right)^{1/2} \exp( -\tau_{c} \xi^2 )
\end{align}
and an exponentially decaying correlation function
\begin{align}
C(t) = \langle \xi(t) \xi(0) \rangle = \frac{1}{2 \tau_{c}} \exp\left( -\frac{|t|}{\tau_{c}} \right).
\end{align}
Thus, the characteristic decay time of the noise correlation is $\tau_{c}$.  In the limit $\tau_{c} \to 0$, $C(t)$ converges to the Dirac delta function $\delta(t)$, so that $\xi(t)$ converges to Gaussian white noise of unit intensity. 
The power spectrum of $\xi(t)$ and its Hilbert transform are given by
\begin{align}
I(\Omega) = \frac{1}{1 + ( \Omega \tau_{c} )^{2}},
\;\;\;
\hat{I}(\Omega) = \frac{\Omega \tau_{c}}{1 + ( \Omega \tau_{c} )^{2}}.
\label{OUPpow}
\end{align}

\begin{figure}[tbp]
  \begin{center}
    \includegraphics[width=0.75\hsize,clip]{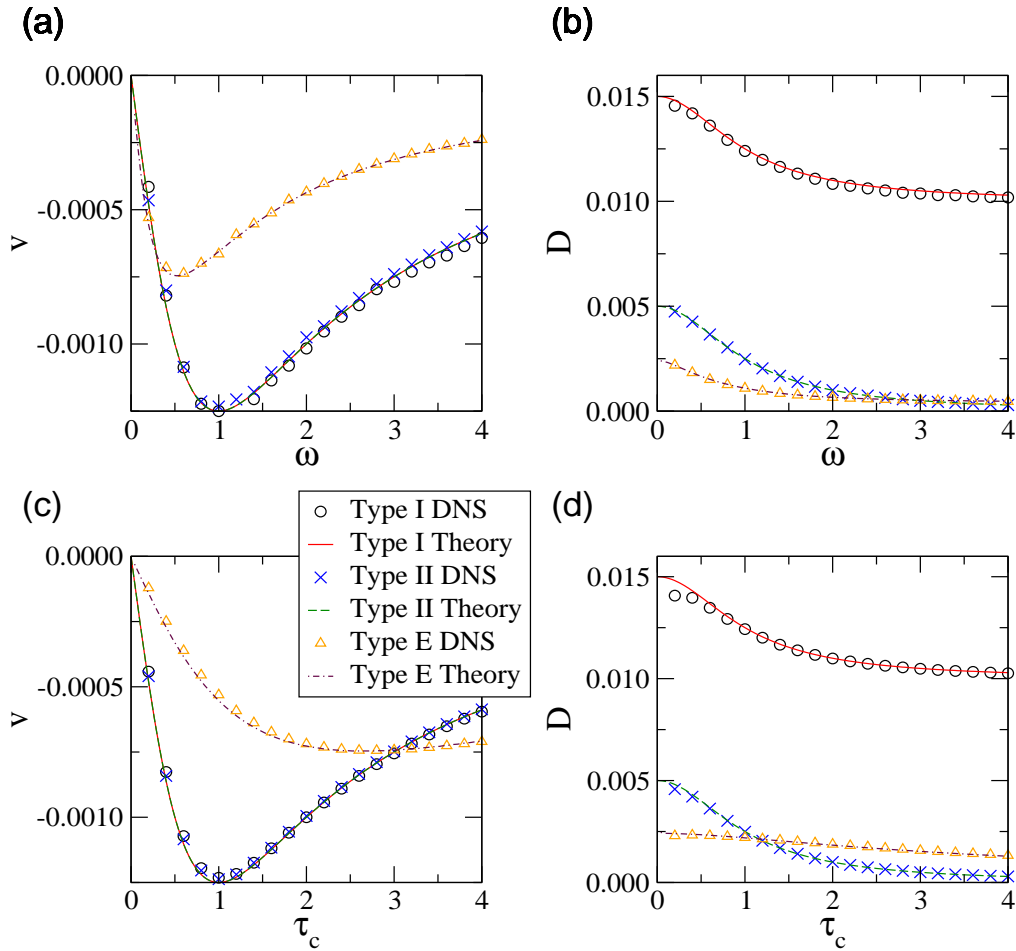}
    \caption{Effective drift and diffusion coefficients $v$ and $D$, respectively, plotted against oscillator frequency $\omega$ [(a) and (b)] and correlation time $\tau_{c}$ [(c) and (d)] for Type-I, II, and E phase sensitivity functions $Z_{I}(\phi)$, $Z_{II}(\phi)$, and $Z_{E}(\phi)$, respectively, driven by Ornstein-Uhlenbeck noise.  Noise intensity $\epsilon = 0.1$ in all cases.  Noise correlation time $\tau_{c} = 1.0$ in (a) and (b), and oscillator frequency $\omega = 1.0$ in (c) and (d).   Data obtained by direct numerical simulations are compared with theoretical curves.}
    \label{OUP}
  \end{center}
\end{figure}

By inserting Eq.~(\ref{OUPpow}) into Eqs.~(\ref{Effv}) and~(\ref{EffD}), the drift and diffusion coefficients are expressed as
\begin{align}
v = - \frac{\epsilon^2}{2} \sum_{\ell=-\infty}^{\infty} |\tilde{Z}_{\ell}|^{2} \frac{ \omega \tau_{c} \ell^{2} }{ 1 + ( \omega \tau_{c} \ell  )^2 },
\end{align}
and
\begin{align}
D = \epsilon^2 \sum_{\ell=-\infty}^{\infty} |\tilde{Z}_{\ell}|^{2} \frac{ 1 }{ 1 + ( \omega \tau_{c} \ell  )^2 }.
\end{align}
Note that $v$ is always non-positive and vanishes in the white-noise limit ($\tau_{c} \to 0$).  That is, the OU noise $\xi(t)$ always tends to slow down the oscillator for arbitrary (smooth) phase sensitivity functions even if $\langle \xi(t) \rangle = 0$ holds on average, which agrees with the result previously obtained by G\'alan~\cite{Galan3}  (however, \cite{Galan3} uses a different definition of the OU process).
Similarly, the diffusion coefficient $D$ is maximized in the white-noise limit.

For the Type-I phase sensitivity function $Z_{I}(\phi)$, $v$ and $D$ are explicitly calculated as
\begin{align}
v = - \frac{\epsilon^2}{4} \frac{ \omega \tau_{c} }{ 1 + ( \omega \tau_{c} )^2 },
\;\;\;
D = \frac{\epsilon^2}{2} \left( 2 + \frac{ 1 }{ 1 + ( \omega \tau_{c})^2 } \right),
    \label{vDoup1}
\end{align}
and for the Type-II $Z_{II}(\phi)$ as
\begin{align}
v = - \frac{\epsilon^2}{4} \frac{ \omega \tau_{c} }{ 1 + ( \omega \tau_{c} )^2 },
  \;\;\;
D = \frac{\epsilon^2}{2} \frac{ 1 }{ 1 + ( \omega \tau_{c} )^2 }.
	\label{vDoup2}
\end{align}
Note that $v$ is the same for both $Z_{I}(\phi)$ and $Z_{II}(\phi)$, whereas $D$ for $Z_{I}(\phi)$ is larger than that for $Z_{II}(\phi)$.  This can easily be seen from the Fourier representations; the only difference between $Z_{I}(\phi)$ and $Z_{II}(\phi)$ is that $Z_{I}(\phi)$ has a non-vanishing constant component $\tilde{Z}_{0} = 1/2$.  For the Type-E phase sensitivity $Z_{E}(\phi)$, we numerically integrate Eqs.~(\ref{Effv}) and (\ref{EffD}) to obtain $v$ and $D$.

In numerical simulations, the noise correlation time is fixed at $\tau_{c} = 1$ and the noise intensity at $\epsilon = 0.1$.  Figures~\ref{OUP}(a) and (b) plot $v$ and $D$ as functions of the oscillator frequency $\omega$ for the three types of phase sensitivity (averaged over 200,000 realizations) and compare them with theoretical values, indicating good agreement.  The drift coefficient $v$ for $Z_{I}(\phi)$ and $Z_{II}(\phi)$ coincide with each other and are minimized at $\omega = \tau_{c}^{-1} = 1$.  The diffusion coefficient $D$ for $Z_{I}(\phi)$ and $Z_{II}(\phi)$ differ from each other and decrease monotonically with $\tau_{c}$.  In particular, $D$ for $Z_{II}(\phi)$ (more generally for $Z(\phi)$ without a constant component $\tilde{Z}_{0}$) tends to vanish at large $\omega$, indicating that the long-time phase diffusion of Type-II oscillators can be very small when the oscillator frequency is large.  The numerical values and theoretical values of $v$ and $D$ for $Z_{E}(\phi)$ are also in good agreement.

\subsection{Noise generated by a damped noisy harmonic oscillator}

Next, we consider colored noise $\xi(t)$ generated by a damped noisy harmonic oscillator (hereafter referred to as DNHO noise),
\begin{align}
  \dot{x} = \omega_0 y - \gamma x + \gamma \eta_x(t),
  \;\;
  \dot{y} = - \omega_0 x - \gamma y + \gamma \eta_y(t),  
\end{align}
where $\eta_x(t)$ and $\eta_y(t)$ are mutually independent Gaussian white noise satisfying $ \langle \eta_x(t) \eta_x(s) \rangle = \langle \eta_y(t) \eta_y(s) \rangle = \delta(t-s)$ and $\langle \eta_x(t) \eta_y(s) \rangle = 0$.  The parameter $\omega_0$ is the frequency of the harmonic oscillations and $\gamma$ is the damping constant.  This process yields two-component noise with a Gaussian stationary PDF,
\begin{align}
  P(x, y) = \frac{1}{\pi \gamma} \exp \left[ - \frac{1}{\gamma} ( x^2 + y^2 ) \right],
\end{align}
and a correlation function $C(t)$ of $x(t)$ with oscillatory decay,
\begin{align}
  C(t) = \langle x(t) x(0) \rangle = \frac{\gamma}{2} e^{- \gamma |t|} \cos( \omega_0 t ).
\end{align}
Thus, the correlation time is given by $\tau_{c} = \gamma^{-1}$.  We use this $x(t)$ as the noise $\xi(t)$ given to the oscillator.  The power spectrum of $x(t)$ and its Hilbert transform are respectively given by
\begin{align}
  I(\Omega) = \frac{\gamma^{2}}{2} \left( \frac{1}{ \gamma^{2} + ( \Omega + \omega_{0} )^{2} } + \frac{1}{ \gamma^{2} + ( \Omega - \omega_{0} )^{2} } \right),
\end{align}
and
\begin{align}
  \hat{I}(\Omega) = \frac{\gamma}{2} \left( \frac{\Omega + \omega_{0} }{ \gamma^{2} + ( \Omega + \omega_{0} )^{2} } + \frac{ \Omega - \omega_{0 }}{ \gamma^{2} + ( \Omega - \omega_{0} )^{2} } \right).
\end{align}

\begin{figure}[tbp]
  \begin{center}
    \includegraphics[width=0.75\hsize,clip]{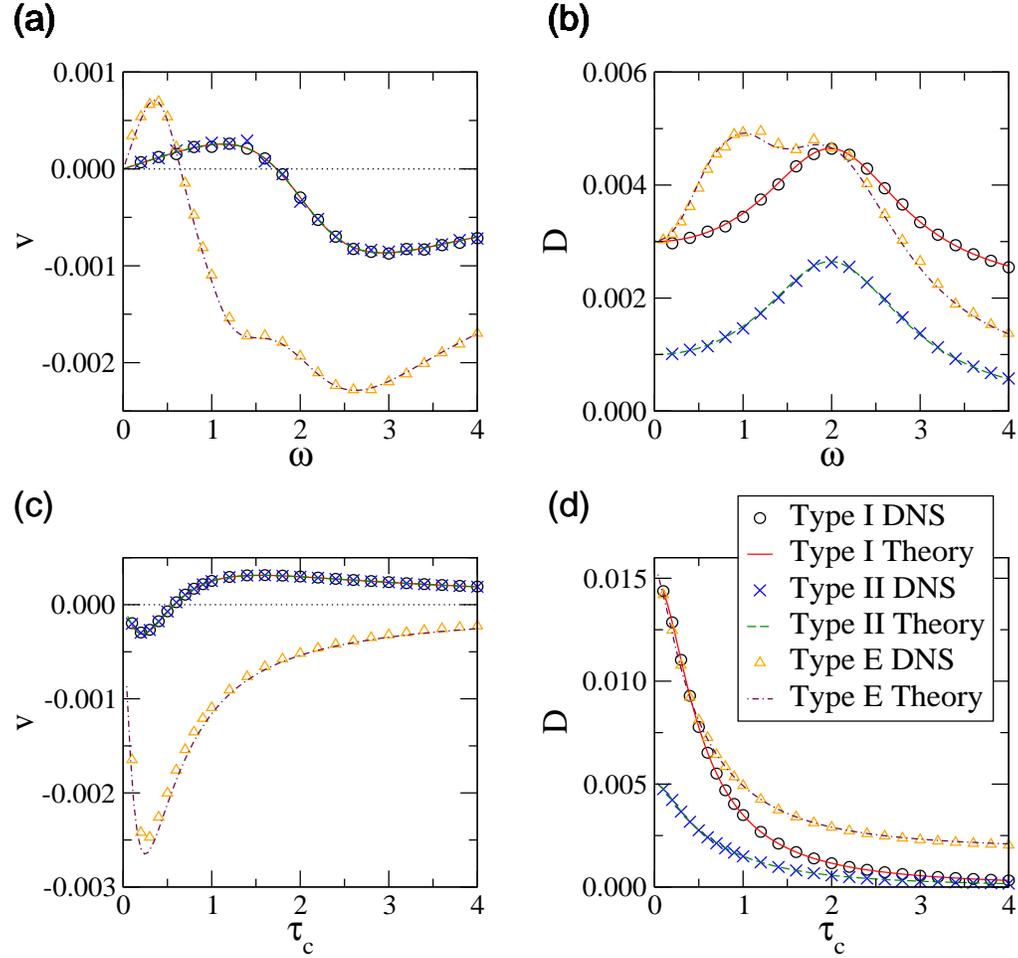}
    \caption{Effective drift and diffusion coefficients, $v$ and $D$ respectively, plotted against oscillator frequency $\omega$ and correlation time $\tau_{c}$ for phase oscillators with Type-I, II, and E phase sensitivity functions $Z_{I}(\phi)$, $Z_{II}(\phi)$, and $Z_{E}(\phi)$, respectively, driven by the damped noisy harmonic oscillator noise.  Noise intensity$\epsilon = 0.1$ and noise frequency $\omega_0 = 2.0$ in all cases. $\tau_{c} = 1.0$ in (a) and (b), and $\omega = 1.0$ in (c) and (d). Data obtained by direct numerical simulations are compared with theoretical curves.}
    \label{DNHO}
  \end{center}
\end{figure}

Effective drift and diffusion coefficients $v$ and $D$ can be analytically calculated for the Type-I phase sensitivity $Z_{I}(\phi)$ as
\begin{align}
  v &= - \frac{\epsilon^2  \gamma}{8} \left( \frac{\omega - \omega_0}{ \gamma^{2} + ( \omega - \omega_0 )^{2} } + \frac{\omega + \omega_0}{ \gamma^{2} + ( \omega + \omega_0 )^{2} } \right),
  \cr \cr
  D &= \frac{\epsilon^{2}  \gamma^2}{4} \left( \frac{1}{\gamma^{2} + (\omega - \omega_0)^{2}} + \frac{4}{\gamma^{2} + \omega_0^{2}} + \frac{1}{\gamma^{2} + (\omega + \omega_0)^{2}} \right),
\end{align}
and for the Type-II phase sensitivity $Z_{2}(\phi)$ as
\begin{align}
  v &= - \frac{\epsilon^2  \gamma}{8} \left( \frac{\omega - \omega_0}{ \gamma^{2} + ( \omega - \omega_0 )^{2} } + \frac{\omega + \omega_0}{ \gamma^{2} + ( \omega + \omega_0 )^{2} } \right),
  \cr \cr
  D &= \frac{\epsilon^{2}  \gamma^2}{4} \left( \frac{1}{\gamma^{2} + (\omega - \omega_0)^{2}} + \frac{1}{\gamma^{2} + (\omega + \omega_0)^{2}} \right).
\end{align}
In the $\omega_0 \to 0$ limit, DNHO noise returns to the OU noise, so that $v$ and $D$ converge to the corresponding results for the OU noise.
Note that values of $v$ coincide again between  $Z_{I}(\phi)$ and $Z_{II}(\phi)$, whereas those of $D$ differ between the two cases.
Values of $v$ and $D$ for the Type-E phase sensitivity $Z_{E}(\phi)$ are calculated by numerically integrating Eqs.(\ref{Effv}) and (\ref{EffD}). 

Figure~\ref{DNHO} plots $v$ and $D$ obtained by direct numerical simulations of Eq.~(\ref{Slow}) (averaged over $200,000$ realizations), and the data are compared with the theoretical results.  The parameters $\epsilon=0.1$ and $\omega_{0}=2$ are fixed and the oscillator frequency $\omega$ or the noise correlation time $\tau_{c} = \gamma^{-1}$ is varied.
In Figs.~\ref{DNHO}(a) and (b) their dependence on $\omega$ with fixed $\tau_{c} = 1$ is shown.  In contrast to the OU case, $v$ can take positive and negative values for all $Z(\phi)$.  $D$ does not monotonically decrease but exhibits a peak (at $\omega = \omega_0$ for $Z_{I}(\phi)$ and $Z_{II}(\phi)$) implying some type of resonance effect.  $D$ for $Z_{I}(\phi)$ is again larger than that for $Z_{II}(\phi)$.
Figures~\ref{DNHO}(c) and (d) show the dependence of $v$ and $D$ on the noise correlation time $\tau_{c}$ with fixed oscillator frequency $\omega = 1$.  The drift coefficient $v$ can take positive values for $Z_{I}(\phi)$ and $Z_{II}(\phi)$.  $D$ decreases monotonically for all types of the phase sensitivity.
In all cases, numerical and theoretical results are in agreement.

\subsection{Noise generated by a chaotic Lorenz model}

\begin{figure}[tbp]
  \begin{center}
    \includegraphics[width=0.75\hsize,clip]{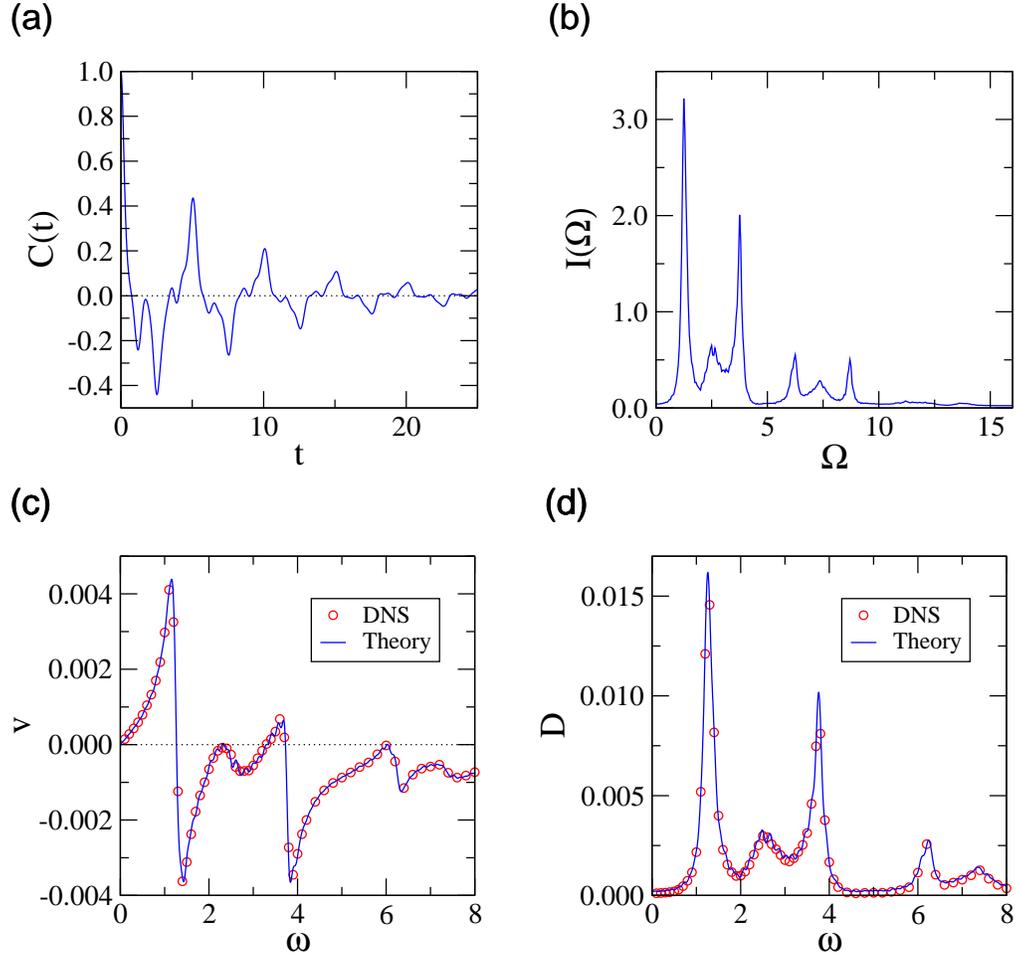}
    \caption{(a) Correlation function $C(t)$ and (b) power spectrum
      $I(\omega)$ of the normalized $x$ variable of the Lorenz model. (c) Drift and (d) diffusion coefficients of the phase with Type-II phase sensitivity $Z_{II}(\phi)$ driven by the Lorenz model.}
    \label{FigLorenz}
  \end{center}
\end{figure}

The Lorenz model~\cite{Strogatz}
\begin{align}
  \dot{x} = p ( -x + y), \;\;\;
  \dot{y} = - x z + q x - y, \;\;\;
  \dot{z} = x y - r z
\end{align}
generates a typical chaotic time sequence.  We use parameter values $p=10$, $q=28$, and $r=0.9$ and apply the normalized time sequence of $x(t)$, 
\begin{align}
  \tilde{x}(t) = \frac{ x(t) - \langle x \rangle }{ \sqrt{ \langle x^2 \rangle } },
\label{normalize}
\end{align}
to the phase model as the colored noise $\xi(t)$, where $\langle \cdots \rangle$ denotes the long-time average.
Figures~\ref{FigLorenz}(a) and (b) show the correlation function and power spectrum of the noise, respectively.  The correlation function exhibits oscillatory decay with several characteristic frequencies, which appear in the power spectrum of the noise as sharp peaks. 

For simplicity, we consider only the Type-II phase sensitivity $Z_{II}(\phi)$.  Figure~\ref{FigLorenz}(c) and (d) compare the drift and diffusion coefficients $v$ and $D$ obtained by direct numerical simulations of Eq.~(\ref{Slow}) (averaged over 100,000 realizations) with the Lorenz model and the theoretical values calculated from the correlation function $C(t)$, which are in agreement.
The noise intensity is fixed at $\epsilon=0.1$ and the oscillator frequency $\omega$ is varied.  The drift and diffusion coefficients $v$ and $D$ show interesting peculiar dependence on $\omega$.  As $\omega$ increases, $v$ increases rapidly and then suddenly decreases, and this is repeated several times.  $D$ exhibits a few sharp peaks, indicating that phase diffusion due to the Lorenz noise can be strongly enhanced at some particular values of the frequency $\omega$.

These results, in particular the behavior of $D$, can easily be understood from the Fourier representation, Eq.~(\ref{D_ft}).  Since $Z_{II}(\phi)$ has only the first harmonic component, $\tilde{Z}_{\pm 1} = \pm i / 2$, Eq.~(\ref{D_ft}) simply gives $D = \epsilon^{2} [ I(\omega) + I(-\omega) ] / 4 = \epsilon^{2} I(\omega) / 2$, namely, $D$ is simply proportional to the power spectrum itself.  In fact, we can see that the curves in Figs.~\ref{FigLorenz}(c) and (d) are identical except for the scaling factor $\epsilon^{2} / 2$.  Moreover, from Eq.~(\ref{v_ft}), we can see that the sudden rise and fall of $v$ is due to the Hilbert transform $\hat{I}(\Omega)$ of the power spectrum $I(\Omega)$ near its sharp peaks, which gives a contribution $1 / (\Omega - \Omega')$ if the peak is approximated by a Dirac $\delta$ function $\delta( \Omega - \Omega' )$. 

\subsection{Noise generated by a chaotic R\"ossler oscillator}

\begin{figure}[tbp]
  \begin{center}
    \includegraphics[width=0.75\hsize,clip]{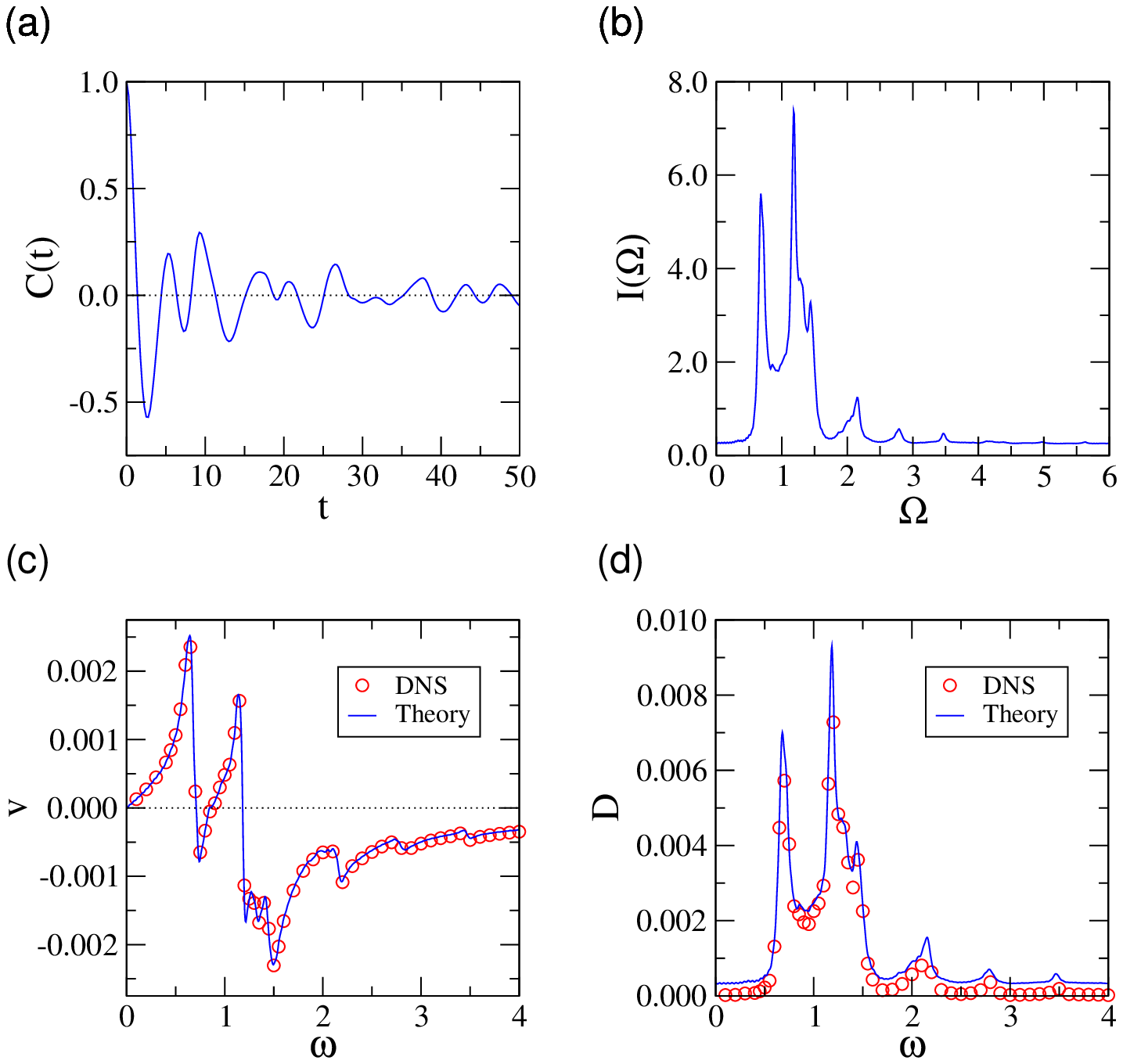}
    \caption{(a) Correlation function $C(t)$ and (b) power spectrum $I(\omega)$ of the normalized $x$ variable of the R\"ossler oscillator. (c) Drift and (d) diffusion coefficients of a phase oscillator with Type-II phase sensitivity $Z_{II}(\phi)$ driven by the R\"ossler model.}
    \label{FigRossler}
  \end{center}
\end{figure}

The R\"ossler oscillator~\cite{Strogatz}
\begin{align}
  \dot{x} = -y - z, \;\;\;
  \dot{y} = - x + a y, \;\;\;
  \dot{z} = b + x z - c z
\end{align}
is another typical example of low-dimensional chaos.  We fix the parameter values at $a=0.3$, $b=0.2$, and $c=5.7$, where the R\"ossler oscillator possesses a ``funnel'' attractor.
The $x$-component is normalized as in Eq.~(\ref{normalize}) and applied to Eq.~(\ref{Slow}) as the noise $\xi(t)$.
Figures~\ref{FigRossler}(a) and (b) show the correlation function and power spectrum, exhibiting oscillatory decay and sharp peaks similar to the Lorenz model.

We consider only the Type-II phase sensitivity $Z_{II}(\phi)$ again.  Figure~\ref{FigRossler}(c) and (d) compare the drift and diffusion coefficients $v$ and $D$ obtained by direct numerical simulations of Eq.~(\ref{Slow}) (averaged over 100,000 realizations) with those calculated from the correlation function.  Noise intensity $\epsilon=0.05$ is fixed and oscillator frequency $\omega$ is varied.  Similarly to the case of the Lorenz model, $v$ and $D$ indicate interesting complex dependence on the oscillator frequency $\omega$, reflecting the peculiar power spectrum of the R\"ossler oscillator.  There is again a good agreement between the numerical and theoretical values.

\section{Summary}

We derived an effective white-noise Langevin equation that describes the long-time phase dynamics of a limit-cycle oscillator driven by general non-Gaussian colored noise.  Effective drift and diffusion coefficients were calculated from the phase sensitivity of the oscillator and the correlation function of the noise.  The results were verified using several types of colored noise sources, i.e., the Ornstein-Uhlenbeck process, the damped noisy harmonic oscillator, and the chaotic Lorenz and R\"ossler models.

Our analysis gave general expressions for drift and diffusion coefficients, applicable to general limit-cycle oscillators driven by arbitrary weak smooth noise.
In a previous study~\cite{Galan3}, G\'alan calculated the frequency shifts of limit-cycle oscillators (the effective drift coefficient $v$ in our notation) driven by colored Ornstein-Uhlenbeck noise, and pointed out that the frequency shift is always negative for arbitrary phase sensitivity.  In contrast, for other types of noise, frequency shifts can also be positive, so that the noise may increase the frequency of the driven oscillator.  We can also calculate the effective diffusion coefficient $D$, which directly reflects the power spectrum of the driving noise.  In particular, for chaotic noises, $D$ exhibited sharp peaks, indicating that the phase diffusion can be greatly enhanced for peculiar frequencies of the limit-cycle oscillator.

The effective white-noise Langevin description enables us to use the powerful classical methods for stochastic processes~\cite{Risken,Gardiner} and thus provides a general framework for analyzing the long-time behavior of limit-cycle oscillators subjected to noise.
Important future topics will include generalization of the present results to multi-dimensional situations and incorporation of deterministic external forcing (e.g., periodic) and mutual interactions.  It is expected that the combined effect of colored noise and other external perturbations or mutual interactions may lead to qualitatively new dynamics.

\acknowledgements{
We gratefully acknowledge Prof. G. Bard Ermentrout for his useful and stimulating discussions.  H.N. and J.-N.T. thank MEXT, Japan (Grant no.\ 22684020 and 20700304). D.S.G. acknowledges the joint support from CRDF (Grant no.\ Y5--P--09--01) and MESRF (Grant no.\ 2.2.2.3/8038).}

\section{Appendix}

\subsection{Derivation of Eq.~(\ref{M1}) from Eq.~(\ref{M10})}

We set $\tau = n T$ with $n$ being an integer.  The right-hand side of Eq.~(\ref{M10}) can be rewritten as
\begin{align}
  &
  \int_{t}^{t + \tau} dt_{1} \int_{t}^{t_{1}} dt_{2} Z'(\omega t_{1} + \psi(t)) Z(\omega t_{2} + \psi(t)) C(t_{1} - t_{2}) \cr
  &=
  \int_{t}^{t+\tau} dt_{1} \int_{0}^{t_{1}-t} ds
  Z'(\omega t_{1} + \psi(t)) Z(\omega (t_{1}-s) + \psi(t)) C(s) \cr
  &\simeq
  \int_{0}^{\infty} ds C(s) \int_{t}^{t + \tau} dt_{1} 
  Z'(\omega t_{1} + \psi(t)) Z(\omega (t_{1}-s) + \psi(t)) \cr
  &=
  \frac{\tau}{2\pi} \int_{0}^{\infty} ds C(s) \int_{0}^{2\pi} d\theta 
  Z'(\theta) Z(\theta - \omega s),
\end{align}
where we have used
\begin{align}
  &
  \int_{t}^{t + \tau} dt_{1} 
  Z'(\omega t_{1} + \psi(t)) Z(\omega (t_{1}-s) + \psi(t)) \cr
  &=
  \sum_{j=0}^{n-1} \int_{0}^{T} dt_{1} 
  Z'(\omega (t_{1}+ t + j T) + \psi(t)) Z(\omega (t_{1} - s + t + j T) + \psi(t)) \cr
  &=
  \sum_{j=0}^{n-1} \frac{T}{2\pi} \int_{0}^{2\pi} d\theta 
  Z'(\theta + \omega t + 2 \pi j + \psi(t)) Z(\theta - \omega s + \omega t + 2 \pi j T) + \psi(t)) \cr
  &=
  \frac{n T}{2\pi} \int_{0}^{2\pi} d\theta 
  Z'(\theta) Z(\theta - \omega s)
  =
  \frac{\tau}{2\pi} \int_{0}^{2\pi} d\theta 
  Z'(\theta) Z(\theta - \omega s).
\end{align}
Substituting these results into Eq.~(\ref{M10}) yields Eq.~(\ref{M1}).

\subsection{Derivation of Eq.~(\ref{M2}) from Eq.~(\ref{M20})}

Setting $\tau = n T$, the right-hand side of Eq.~(\ref{M20}) can be transformed as
\begin{align}
  &\int_{t}^{t + \tau} dt_{1} \int_{t}^{t + \tau} dt_{2} Z(\omega t_{1} + \psi(t)) Z(\omega t_{2} +
  \psi(t)) C(t_{1} - t_{2}) \cr
  &=
  \int_{t}^{t + \tau} dt_{1} \int_{t_{1} - t - \tau}^{t_{1}-t} ds
  Z(\omega t_{1} + \psi(t)) Z(\omega (t_{1}-s) + \psi(t)) C(s) \cr
  &\simeq
  \int_{-\infty}^{\infty} ds C(s) \int_{t}^{t+\tau} dt_{1} 
  Z(\omega t_{1} + \psi(t)) Z(\omega (t_{1}-s) + \psi(t)) \cr
  &=
  \frac{\tau}{2\pi} \int_{-\infty}^{\infty} ds C(s) \int_{0}^{2\pi} d\theta Z(\theta) Z(\theta - \omega s),
\end{align}
where we have approximated the range of the integral of the correlation function $C(s)$ over $[t_{1}-t-\tau, t_{1}-t]$ as $[-\infty, +\infty]$ by assuming that the decay time of $C(s)$ is much shorter than $\tau$ and that $t_{1} - t - \tau < 0$ and $t_{1} - t > 0$ hold.
In deriving the final expression, we used
\begin{align}
  &
  \int_{t}^{t+\tau} dt_{1} Z(\omega t_{1} + \psi(t)) Z(\omega (t_{1}-s) + \psi(t)) \cr
  &=
  \sum_{j=0}^{n-1} \int_{0}^{T} dt_{1} 
  Z(\omega (t_{1} + t + j T) + \psi(t)) Z(\omega (t_{1} - s + t + j T) + \psi(t)) \cr
  &=
  \sum_{j=0}^{n-1} \frac{T}{2\pi} \int_{0}^{2\pi} d\theta 
  Z(\theta + \omega t + 2 \pi j + \psi(t)) Z(\theta - \omega s + \omega t + 2 \pi j + \psi(t)) \cr
  &=
  \sum_{j=0}^{n-1} \frac{T}{2\pi} \int_{0}^{2\pi} d\theta 
  Z(\theta) Z(\theta - \omega s) 
  =
  \frac{n T}{2\pi} \int_{0}^{2\pi} d\theta 
  Z(\theta) Z(\theta - \omega s) \cr
  &=
  \frac{\tau}{2\pi} \int_{0}^{2\pi} d\theta 
  Z(\theta) Z(\theta - \omega s).
\end{align}
We obtain Eq.~(\ref{M2}) by substituting the above result into Eq.~(\ref{M20}).

\subsection{The Morris-Lecar model}

The Morris-Lecar model of a spiking neuron is given by the following set of two-variable ordinary differential equations~\cite{Koch,Holmes,Rinzel}:
\begin{align}
C \dot{V}(t) &= g_{Ca} m_{\infty}(V) ( V_{Ca} - V ) + g_{K} ( V_{K} - V ) + g_{L} ( V_{L} - V ) + I, \cr
\dot{w}(t) &= \phi \left( \frac{ w_{\infty}(V) - w }{ \tau_{w}(V) } \right),
\end{align}
where
\begin{align}
m_{\infty}(V) &= \frac{1}{2} \left[ 1 + \tanh \left( \frac{V - V_{1}}{V_{2}} \right) \right], \cr
w_{\infty}(V) &= \frac{1}{2} \left[ 1 + \tanh \left( \frac{V - V_{3}}{V_{4}} \right) \right], \cr
\tau_{w}(V) &= \left[ \cosh \left( \frac{V - V_{3}}{2 V_{4}} \right) \right]^{-1}.
\end{align}
Here, $V$ represents membrane potential and $w$ is an activation variable for potassium.
The parameter values are chosen as $\phi = 0.23$, $g_{L} = 2.0$, $g_{Ca} = 4.0$, $g_{K} = 8.0$, $C = 20.0$, $V_{K} = -84.0$, $V_{L} = -60.0$, $V_{Ca} = 120.0$, $V_{1} = -1.2$, $V_{2} = 18.0$, $V_{3} = 12.0$, $V_{4} = 17.4$, and $I = 37.5$~\cite{Rinzel}.
This model exhibits limit-cycle oscillations via homoclinic bifurcation near $I \simeq 35$.

We set the origin of the phase $\phi = 0$ at the point where $V$ exceeds $0$ from below.  The phase sensitivity function $Z_{E}(\phi)$ for this model can be numerically obtained by the adjoint method as explained in~\cite{Kopell,Holmes}.   It has an exponentially decaying part, which tends to dominate the whole function as the parameter $I$ approaches the bifurcation point. 
The above set of parameter values gives a fixed frequency $\omega = 0.198$.  However, note that we may still set $\omega$ arbitrarily as in Figs.~\ref{OUP} and~\ref{DNHO} by rescaling the time appropriately ($Z_{E}(\phi)$ is not affected by time rescaling).

\end{document}